\documentclass[twocoloumn]{nato}  

\usepackage{txfonts}

\usepackage{makeidx}
\makeindex

\usepackage{sprrefn} 

\usepackage[draft]{hyperref}

\newdisplay{guess}{Conjecture}

\begin{document}

\firstpage{1}

\begin{opening}
\title{Experiments using high-$T_c$ versus low-$T_c$ Josephson contacts}

\runningtitle{Experiments using high-$T_c$/low-$T_c$ Josephson
contacts}

\author{Ariando,$^1$ H.~J.~H.~Smilde,$^1$ C.~J.~M.~Verwijs,$^1$ G.~Rijnders,$^1$
D.~H.~A.~Blank,$^1$ H.~Rogalla,$^1$ J.~R.~Kirtley,$^2$
C.~C.~Tsuei,$^2$ and H. Hilgenkamp$^1$\\
\email{h.hilgenkamp@utwente.nl}}

\runningauthor{Ariando et al.}

\institute{
$^1$Faculty of Science and Technology and MESA$^+$
Institute for Nanotechnology, University of Twente, P.O. Box 217,
7500 AE Enschede, The Netherlands\\
$^2$IBM T. J. Watson Research Center, Yorktown Heights, New York
10598, USA}

\begin{abstract}
Remarkably rich physics is involved in the behavior of hybrid
Josephson junctions, connecting \index{Josephson
contact!high-$T_c$/low-$T_c$} high-$T_c$ and low-$T_c$
superconductors. This relates in particular to the different order
parameter symmetries underlying the formation of the superconducting
states in these materials. Experiments on
\index{high-$T_c$/low-$T_c$ Josephson contact} high-$T_c$/low-$T_c$
contacts have also played a crucial role in settling the decade-long
\index{$d_{x^2-y^2}$-wave pairing symmetry} $d$-wave versus $s$-wave
debate in cuprate superconductors. Recently, such hybrid junctions
have enabled more detailed pairing symmetry tests. Furthermore, with
these junctions, complex arrays of $\pi$-rings have been realized,
enabling studies on spontaneously generated fractional flux quanta
and their mutual interactions. Steps toward novel superconducting
electronic devices are taken, utilizing the phase-shifts inherent to
the $d$-wave superconducting order parameter. This paper is intended
to reflect the current status of experiments using high-$T_c$ and
low-$T_c$ Josephson contacts.
\end{abstract}


\end{opening}

\section{Introduction}


The introduction of additional \index{phase-shift} phase shifting
elements in superconducting loops leads to remarkable effects
\cite{vanharlingenrmp,tsueirmp}. This can be achieved e.g. by using
superconductors with unconventional \index{pairing symmetry} pairing
symmetry \cite{geshkenbein,geshkenbein2,sigristjpl,tsueirmp}, by
incorporating $\pi$-Josephson junctions \cite{bulaevskii}, or by
employing trapped flux quanta \cite{majerapl02} or current injection
\cite{goldobinprl04}. In \index{phase-sensitive experiment}
phase-sensitive tests of the pairing symmetry, superconductors with
unconventional pairing symmetry were used to create a $\pi$
phase-shift in the ring \index{$\pi$-ring} ($\pi$-ring), leading to a
complementary magnetic field dependence of the critical current of the
ring \cite{wollmansquid,wollmanprl95} and to the \index{half-flux
quantum} half-flux quantum effect \cite{tsueiybcoprl94}.

These interesting observations are due to the physics associated
with $d$-wave symmetry taking place at the interface of the
high-$T_c$ cuprate superconductors, as reviewed for example in
\cite{vanharlingenrmp,tsueirmp,Kashiwayarpp00,hilgenkamprmp,tafurirpp05}.
Other interesting phenomena primarily associated with $d$-wave
symmetry include e.g. Andreev bound states
\cite{huprl94,Kashiwayarpp00,Chesca_Condmat0402131,chescaprb06}, the
presence of the second harmonic in the critical current versus phase
relation \cite{golubovrmp}, or vortex splintering \cite{mintsprl89}.

In the phase-sensitive experiments, superconducting loops were
fabricated using \index{high-$T_c$/low-$T_c$ Josephson contact} a
high-$T_c$ cuprate connected to a low-$T_c$ superconductor
\cite{wollmansquid,wollmanprl95}, or using tricrystal or
tetracrystal grain boundary junctions
\cite{tsueiybcoprl94,tsueirmp}. In this article we will concentrate
only on the first route, using a high-$T_c$ material connected to a
low-$T_c$ superconductor.

\begin{figure}[h!]
\centerline{\includegraphics[width=4.8in]{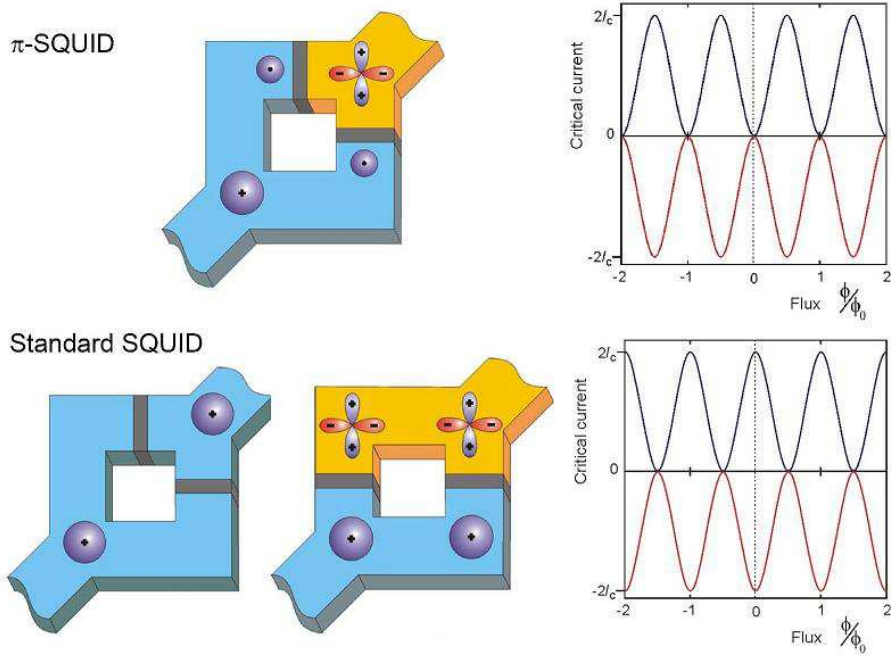}}
\caption[]{(Color) A schematic representation of a $\pi$-ring (top)
and a 0-ring (bottom), and the expected magnetic field dependencies
of their critical currents as shown at the right-hand side.}
\label{pisquid}
\end{figure}

A schematic of a $\pi$-ring ($\pi$-SQUID) and
0-ring (standard SQUID) employing \index{high-$T_c$/low-$T_c$
Josephson contact} high-$T_c$ and low-$T_c$ Josephson contacts is
depicted in Fig. \ref{pisquid}. In the limit of a
\index{low-inductance SQUID} low inductance $L\ll\Phi_0/I_c$, with
$\Phi_0=h/2e=2.07\times10^{-15}$~Wb the flux quantum and $I_c$
the critical current of the junctions in the ring, the $\pi$-SQUIDs have
$I_c(B)$ characteristics complementary to those of standard SQUIDs.
For $L\gg\Phi_0/I_c$, the energetic ground state involves a
spontaneously generated flux of $\frac{1}{2}\Phi_0$ in the SQUID
loop. To compensate for the built-in $\pi$-phase shift
\index{phase-shift} a spontaneous circulating current flows in the
ring, in either the clock- or counter-clockwise direction. The
magnetic flux associated with this persistent circulating current is
a fraction of a flux quantum, growing asymptotically to a
\index{half-flux quantum} half-flux quantum in the large inductance
limit \cite{kirtleyprb}. In the limiting case when the inner
diameter of the $\pi$-ring in Fig. \ref{squid} is reduced to zero,
the structure represents the corner junction \cite{wollmanprl95}. A
corner junction behaves similarly to a $\pi$-ring. In this structure,
the  \index{$d_{x^2-y^2}$-wave pairing symmetry}
$d$-wave order parameter of the high-$T_c$ cuprate induces
a difference of $\pi$ in the Josephson phase shift $\Delta\phi$
between the two junctions (facets). For facet
lengths $a$ in the small limit, that is, $a\ll\lambda_J$, the corner
junction has $I_c(B)$ characteristics with zero critical current in
the absence of magnetic fields, in stark contrast to the Fraunhofer
pattern for a uniform junction. For facet lengths $a$ in the wide
limit, that is, $a\gg\lambda_J$, the lowest-energy ground state of
the system is expected to be characterized by the spontaneous
generation of a half-integer flux quantum
at the corner. This half-fluxon  provides a
further $\pi$-phase change between neighbouring
facets, either adding or subtracting to the $d$-wave induced
\index{phase-shift} $\pi$-phase shift, depending on the
half-flux-quantum polarity. In both cases,
this leads to a lowering of the Josephson coupling energy across the
barrier, as this energy is proportional to $(1-\cos\Delta\phi)$.

\section{Preparation of high-$T_c$ and low-$T_c$ ramp-type Josephson contacts}

Various early attempts have been made to prepare Josephson contacts
between \index{high-$T_c$/low-$T_c$ Josephson contact} high-$T_c$
and low-$T_c$ superconductors, such as YBa$_2$Cu$_3$O$_7$ (YBCO) and
Nb
\cite{akohjjap88,akohieee89,akohapl90,fujimakijjap90,fujimakiieee91,huntapl90,huntieee91,footeieee91,teraijjap93,wollmansquid,brawnerprb94,brawnerphysicac94,mathaiprl95,gimjdp96,wollmanprl95,teraiieee95,brawnerprb96,usagawaapl98}.
Oxygen migration due to the chemical reactivity of Nb with oxygen,
combined with the sensitivity of YBCO to oxygen loss, presents a
difficulty for the preparation of a good electrical connection
between both superconductors. Another crucial step, especially using
the ramp type configuration, is the structuring of the
superconducting base electrode. Unfortunately, this procedure can
severely degrade the quality of the base electrode near the
interface. Transmission electron microscopy (TEM) studies of YBCO/Au
ramp-type interfaces \cite{wenapl99} clearly show an amorphous layer
with a thickness up to 2~nm at the ramp edge between the high-$T_c$
base electrode and the Au layer deposited at the freshly milled ramp
edge.

Using the ramp-type interface configuration
\cite{gaophysicac,gaoapl,verhoevenapl}, we have been able to
establish a fabrication procedure \cite{smildeapl} for all-thin-film
Josephson junctions, in which controllably a high-$T_c$ cuprate
superconductor (such as YBCO or Nd$_{2-x}$Ce$_x$CuO$_4$ (NCCO)) is
connected with a low-$T_c$ material (Nb) along freely chosen
directions of the high-$T_c$ cuprates. A thin \index{interlayer}
interlayer is incorporated in the junctions to obtain an increased
transparency. The interlayer restores the surface damaged by ion
milling and has the advantage of an in-situ barrier deposition
between the two superconductors, leading to clean and well-defined
interfaces. To avoid oxygen migration at the interface, a thin but
chemically closed barrier separating the materials is used, for
which Au is found to be the most suitable
material~\cite{smildephysicac,smildeieee}.

\begin{figure}[h!]
\centerline{\includegraphics[width=4.4in]{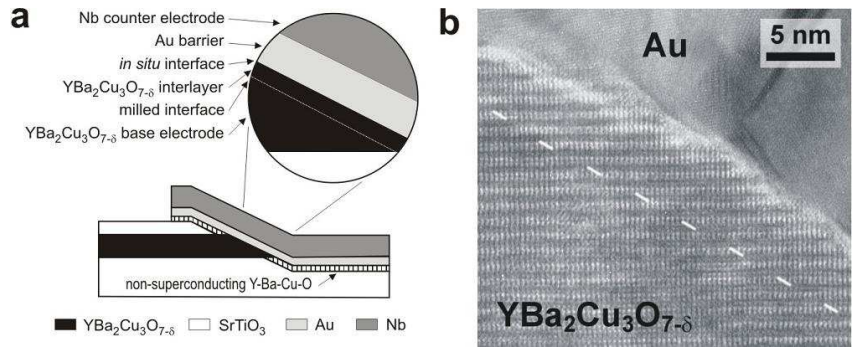}}
\caption[]{(a)~Schematic cross section of the
YBa$_2$Cu$_3$O$_{7-\delta}$/Au/Nb ramp-type junction including the
interlayer. (b)~Bright-field transmission electron microscopy image
of the YBa$_2$Cu$_3$O$_{7-\delta}$/Au interface at the ramp-edge
area, including an interlayer of 6~nm deposited
YBa$_2$Cu$_3$O$_{7-\delta}$. Crystalline YBa$_2$Cu$_3$O$_{7-\delta}$
material is observed up to the Au interface, while no clear
interface is observed between the base electrode and the interlayer
(dashed line) [from \cite{smildeapl}].} \label{interlayer}
\end{figure}

For the preparation, first a [001]-oriented high-$T_c$ cuprate and a
SrTiO$_3$ insulator layer are epitaxially grown by pulsed laser
deposition (PLD) on [001]-oriented SrTiO$_3$ single crystal
substrates. In these films, beveled edges (ramps) are etched by
Ar-ion milling under an angle of $45^\circ$ using a photoresist
stencil, yielding ramps with an angle of $\sim20^\circ$ with the
substrate plane \cite{blank97rampangle}. In order to facilitate a
good alignment of the junction with the $\langle100\rangle$-axes of
the high-$T_c$ film, edge-aligned substrates are used, with an
alignment accuracy better than $1^\circ$. After stripping of the
photoresist, a low-voltage ion mill step is applied to clean the
surface, in-situ followed by an annealing step and the deposition of
a thin interlayer of $5\--7$~nm YBCO at a condition similar to the
deposition of the YBCO base-electrode. The annealing procedure is
introduced to recrystallize residual amorphous material present at
the ramp edge. This is followed by an in-situ deposition of
Au-barrier at room temperature. After deposition of the Au-barrier
layer with a thickness ranging from 6 to 120~nm, a photoresist
lift-off stencil is applied to define the junction area. Before Nb
deposition, maximally 2~nm of the Au layer is removed by rf-sputter
etching, followed in situ by dc-sputter deposition of 150~nm Nb.
After lift-off, the redundant uncovered Au is removed using ion
milling.

A schematic of the junction obtained in this way is presented in
Fig. \ref{interlayer}a. The \index{phase-shift} interlayer concept
employs the difference in homoepitaxial and heteroepitaxial growth of
high-$T_c$ material. The thin interlayer is anticipated to be
superconducting only if deposited on the YBCO ramp area, whereas on
the SrTiO$_3$ substrate and isolation layer it is anticipated not to
become superconducting \cite{smildeapl}. Figure \ref{interlayer}b
presents a TEM micrograph of the ramp edge area near the YBCO/Au
interface. Because of the application of the thin YBCO interlayer,
crystalline high-$T_c$ material extends up to the Au barrier, and an
amorphous layer was never observed. The interface between the base
electrode and the interlayer could not be distinguished by TEM,
indicating nearly perfect homoepitaxial growth.

Following this procedure results in normal state resistance $R_nA$
values of $\sim10^{-12}~\Omega$m$^2$ at liquid helium temperature.
By adapting the Au-barrier thickness $d_{Au}$, the junction critical
current density can be tuned in a wide range from $10^5$~A/m$^2$ for
$d_{Au}\sim{120}$~nm, up to values approaching $10^9$~A/m$^2$ for
$d_{Au}\sim{7}$~nm. In the following, we will review recent
experiments based on \index{high-$T_c$/low-$T_c$ Josephson contact}
high-$T_c$ versus low-$T_c$ contacts prepared using the procedure
that has been described in this section.

\section{Pairing symmetry test experiments}

Understanding the nature of the ground state and its low-lying
excitations in the copper oxide superconductors is a prerequisite for
determining the origin of high temperature superconductivity. A
superconducting order parameter (that is, the energy gap) with a
predominantly $d_{x^2-y^2}$ symmetry is well-established
\cite{vanharlingenrmp,tsueirmp}. There are, however, several important
issues that remain highly controversial. For example (in hole-doped
compound such as YBCO) various deviations from a pure
\index{$d_{x^2-y^2}$-wave pairing symmetry} $d$-wave pair state, such
as the possibility of Cooper pairing with
\index{broken time-reversal symmetry (BTRS)}
\index{time-reversal symmetry!broken (BTRS)}
broken time-reversal symmetry (BTRS) or an admixed
$d_{x^2-y^2}+s$ pair state, have been theoretically predicted
\cite{laughlinscience88,varmaprl99,sigristptp98,lofwandersst01} and
actively sought in numerous experimental studies
\cite{spielmanprl92,lawrenceprl92,kaminskinature02,simonprl02,fauqucondmat05,convingtonprl97,daganprl01,sharoniprb02,mathaiprl95,schulzapl}.
Furthermore, a transition of the \index{pairing symmetry} pairing
symmetry from \index{$d_{x^2-y^2}$-wave pairing symmetry} $d$-wave
behavior to $s$-wave-like behavior was also suggested as a function of
doping \cite{Skintadtos,Biswas,Qazilbash} and temperature \cite{Balci}
in various \index{electron doped cuprate} electron doped compounds.

In view of this ongoing discussion, there is a need for further
\index{phase-sensitive experiment} phase-sensitive experiments as a
function of doping and temperature, and specifically for studying
the possible existence of BTRS states. Many phase-sensitive
experiments have been performed to look for evidence of BTRS
\cite{vanharlingenrmp,tsueirmp,mathaiprl95,schulzapl}. From these it
has been concluded that, if present, the imaginary component must be
quite small for high-$T_c$ cuprates over a broad range of doping
\cite{tsueiprl04}. Tsuei and Kirtley succeeded in performing
phase-sensitive measurements for various compounds and temperatures
based on grain boundary junctions and \index{half-flux quantum}
half-flux quantum effects \cite{tsueirmp}. Geometrical restrictions
of the grain boundaries makes such experiments very challenging,
especially for investigations as a function of momentum. In this
section we describe various phase-sensitive experiments that have
been performed based on \index{high-$T_c$/low-$T_c$ Josephson
contact} high-$T_c$/Nb Josephson contacts.

\subsection{Low-inductance-SQUID interferometry}

Superconducting quantum interference devices present excellent tools
to perform \index{phase-sensitive experiment} phase-sensitive
experiments on the order parameter symmetry in superconductors
\cite{wollmansquid,brawnerprb94,mathaiprl95,vanharlingenrmp,schulzapl}.
In a dc SQUID in which an isotropic $s$-wave superconductor contacts
the superconductor to be studied in two crystal orientations, the
critical current versus the applied magnetic flux dependence
contains information about the relative phase and magnitude of the
order parameter wave function at both contacts. Previous SQUID
experiments on the order parameter symmetry of the high-$T_c$
cuprates have been performed using, for example, Pb as a counter
electrode. In these experiments, single crystals \cite{brawnerprb94}
as well as twinned and untwinned thin films
\cite{wollmansquid,mathaiprl95} of YBa$_2$Cu$_3$O$_7$ have been
used. These SQUIDs generally had inductances of $LI_c\approx\Phi_0$
or larger.

\begin{figure}[h!]
\centerline{\includegraphics[width=3.4in]{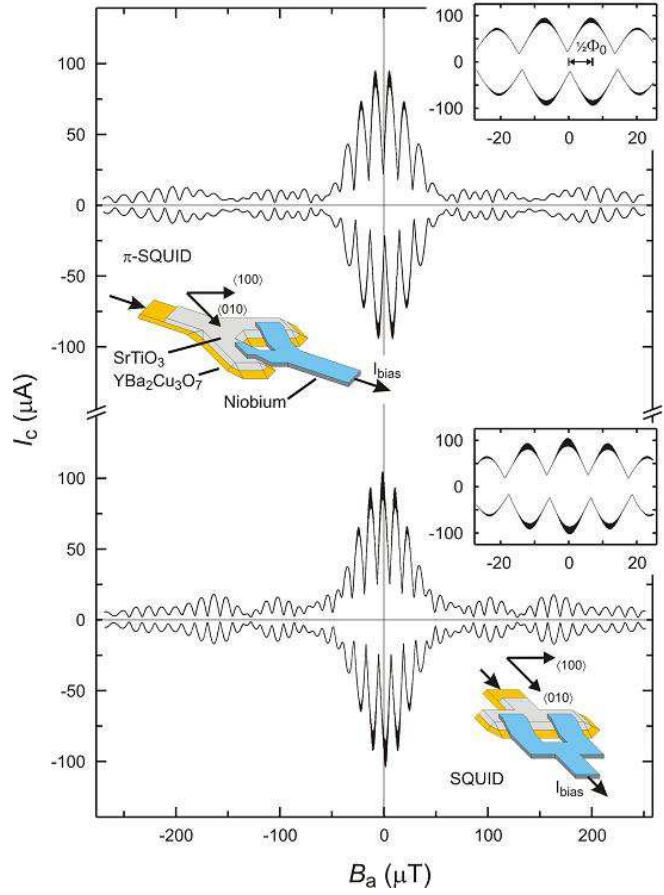}}
\caption[]{(Color) Critical current as a function of the applied
magnetic field at $T=4.2$~K of a $\pi$-SQUID (top) and standard
SQUID (bottom). The insets present schematically the corresponding
configuration and an enlargement of the $I_{c}(B)$ dependence near
zero field. Both SQUIDs are in the low inductance limit [from
\cite{smildeprb}].} \label{squid}
\end{figure}

\index{low-inductance SQUID} Low-inductance SQUIDs enable a more
precise analysis \cite{schulzapl,smildeprb}. Low-inductance
all-high-$T_c$ $\pi$-SQUIDs have been prepared using tetra-crystal
substrates, providing clear evidence for predominant
\index{$d_{x^2-y^2}$-wave pairing symmetry} $d_{x^2-y^2}$-wave order
parameter symmetry \cite{schulzapl}. These dc $\pi$-SQUIDs were
based on symmetric $45^\circ$ [001]-tilt grain-boundary junctions.
By definition, grain-boundary junctions are subject to limitations
with respect to the orientation of the superconductors on both sides
of the junction interface. Additionally, a complicating factor is
presented by the fact that both electrodes are characterized by the
order parameter symmetry under investigation. Junctions combining a
well-characterized isotropic superconductor with the superconductor
to be studied provide the ability to probe the order parameter in
any desired orientation.

We have fabricated \index{low-inductance SQUID} low-inductance
SQUIDs based on \index{high-$T_c$/low-$T_c$ Josephson contact}
high-$T_c$ versus low-$T_c$ Josephson contacts. Sketches of the
devices are shown in the insets of Fig. \ref{squid}. If the
high-$T_c$ order parameter is probed by an isotropic superconductor
along the same main crystal orientation in a dc SQUID (bottom),
i.e., both junctions are oriented in parallel, a maximum critical
current is then observed in the absence of an applied magnetic field
as shown in Fig. \ref{squid} (bottom). When the junctions are
oriented perpendicular with respect to each other in the
[100]-directions, the high-$T_c$ order parameter symmetry induces an
additional \index{phase-shift} phase-shift in the SQUID-loop. A time
reversal invariant $d_{x^2-y^2}$ order parameter symmetry of the
high-$T_c$ cuprate corresponds to a \index{phase-shift} $\pi$
phase-shift, and maxima in the critical current are observed at a
magnetic field equivalent to $\frac{1}{2}\Phi_0$ as depicted in Fig.
\ref{squid} (top). A \index{broken time-reversal symmetry (BTRS)}
time-reversal symmetry breaking (TRSB) order parameter, such as
predominant $d_{x^2-y^2}$ pairing with an imaginary \index{$s$-wave
admixture} $s$-wave admixture, results in a deviation from $\pi$,
and consequently the maximum critical current occurs at an applied
flux differing from $\frac{1}{2}\Phi_0$. Therefore no evidence has
been found for imaginary admixtures. Twinning of the film prevents
the determination of possible real admixtures.

\subsection{Josephson junction modulation}
\label{zigzag}

Multiple \index{$\pi$-ring} $\pi$-loops placed controllably at
arbitrary positions would enable more detailed and systematic
studies of the order parameter symmetry and its effects on Josephson
devices \cite{smildeprl,ariandoprl}, as well as the realization of
theoretically proposed elements for superconducting (quantum)
electronics \cite{beasleyieee,ioffenature,blatterprb}.

\begin{figure}[h!]
\centerline{\includegraphics[width=4.8in]{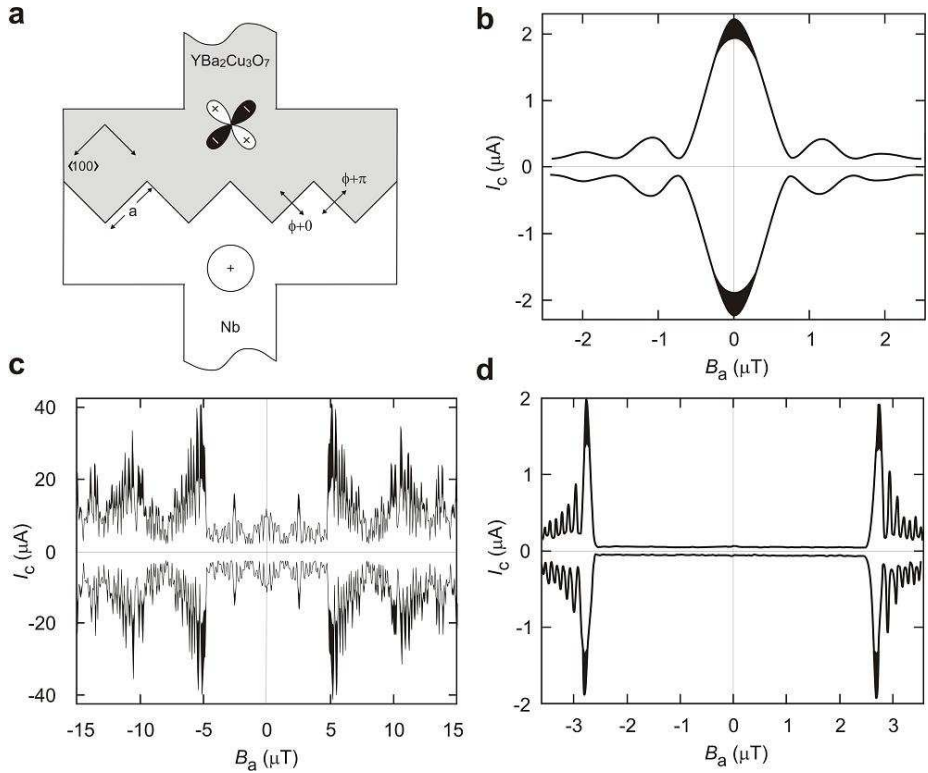}}
\caption[]{(a) A schematic representation of a zigzag junction.
Critical current $I_c$ as a function of applied magnetic field $B_a$
for (b) a straight junction, and (c) a YBCO/Nb and (d) a NCCO/Nb
zigzag array comprised of 80 facets of $5~\mu$m width $(T=4.2$~K)
[from \cite{smildeprl} and \cite{ariandoprl}].} \label{nccoicb}
\end{figure}

We have fabricated well-defined \index{zigzag junction}
\index{junction!zigzag} zigzag-shaped ramp-type Josephson junctions
between various high-$T_c$ superconductors and Nb, and used such
structures to test the order parameter symmetry of these high-$T_c$
compounds. The zigzag configuration is depicted in Fig.
\ref{nccoicb}a. In this structure, all interfaces are aligned along
one of the $\langle100\rangle$ directions of the cuprate, and are
designed to have identical critical current density ($J_c$) values.
If the high-$T_c$ cuprate were an $s$-wave superconductor, there would
be no significant difference between a zigzag and a straight junction,
aligned along one of the facet$^\prime$s direction. With the
high-$T_c$ superconductor having a \index{$d_{x^2-y^2}$-wave pairing
symmetry} $d_{x^2-y^2}$-wave symmetry, the facets oriented in one
direction experience an additional \index{phase-shift} $\pi$-phase
difference compared to those oriented in the other direction. For a
given number of facets, the characteristics of these zigzag structures
then depend on the ratio of the facet length $a$ and the Josephson
penetration depth $\lambda_J$; see, e.g., \cite{zenchukprb69}. In the
small facet limit, $a\ll\lambda_{J}$, the zigzag structure can be
envisaged as a one-dimensional array of Josephson contacts with an
alternating sign of $J_c$, leading to anomalous magnetic field
dependencies of the critical current. In the large facet limit, the
energetic ground state includes the spontaneous formation of
\index{half-flux quantum} half-integer magnetic flux quanta at the
corners of the zigzag structures, as seen in
\cite{hilgenkampnature}. All experiments described in this section are
in the small facet limit.

Figure \ref{nccoicb}c and \ref{nccoicb}d shows the $I_c(B_a)$
dependence for a YBCO/Nb and a NCCO/Nb \index{zigzag junction}
zigzag array with 80 facets having a facet length of $5~\mu$m,
respectively. The $I_c(B_a)$ dependencies of these zigzag structures
clearly exhibit the characteristic features with an absence of a
global maximum at $B_a(0)$ and the sharp increase in the critical
current at a given applied magnetic field, resembling in their basic
features the ones observed for asymmetric $45^\circ$ [001]-tilt
grain boundary junctions. This behavior can only be explained by the
facets being alternatingly biased with and without an additional
$\pi$-phase change
\cite{copettiphysicac,hilgenkampprb86facet,mannhartzphysb,mintsprb}.
This provides a direct evidence for a $\pi$-phase shift in the pair
wave function for orthogonal directions in momentum space and thus
for a predominant $d_{x^2-y^2}$ order parameter symmetry. If the
order parameter were to comprise an imaginary \index{$s$-wave
admixture} $s$-wave admixture, the $I_c(B_a)$ dependencies for the
\index{zigzag junction} zigzag junctions would be expected to
display distinct asymmetries, especially for low fields
\cite{smildeprl}. In addition, the critical current at zero applied
field is expected to increase with the fraction of \index{$s$-wave
admixture} $s$-wave admixture. From the high degree of symmetry of
the measured characteristics of Figs. \ref{nccoicb}c and
\ref{nccoicb}d and the very low zero field $I_c$, an upper limit of
an imaginary \index{$s$-wave admixture} $s$-wave symmetry admixture
to the predominant $d_{x^2-y^2}$ symmetry of $1\%$ for YBCO can be
set and no indication for subdominant symmetry components for NCCO
can be distinguished. The $d$-wave result for the electron-doped
superconductor corroborates the results obtained using grain
boundary junctions \cite{tsueinccoprl,Chesca}

To investigate a possible change of the order parameter symmetry
with doping \cite{Biswas,Qazilbash}, we have fabricated similar
\index{zigzag junction} zigzag structures using
Nd$_{1.835}$Ce$_{0.165}$CuO$_4$/Nb junctions. The results also
indicated a predominant \index{$d_{x^2-y^2}$-wave pairing symmetry}
$d_{x^2-y^2}$-wave symmetry. When cooling the samples to
$T=1.6$~K all the basic features displayed by the structures at
$T=4.2$~K remain unaltered. We thus see no indication for an order
parameter symmetry crossover for Nd$_{1.835}$Ce$_{0.165}$CuO$_4$ in
this temperature range, as was reported for
Pr$_{2-x}$Ce$_x$CuO$_{4-y}$ \cite{Balci}. Similar results were
obtained for optimally doped samples upon cooling to $T=1.6$~K.

\subsection{Angle-resolved electron tunneling}

The upper limit of an imaginary $s$-wave symmetry admixture
\index{$s$-wave admixture} can be determined using the experiments
described above. However, these experiments used geometries in which
the junction normals are perpendicular to the $a$- and $b$- axes.
They are, therefore, insensitive to an
imaginary $d_{xy}$ component which has nodes along the $a$- and
$b$-axes. Another possible modification of the pure
\index{$d_{x^2-y^2}$-wave pairing symmetry} $d$-wave gap parameter
is an admixed $d_{x^2-y^2}+s$ pair state, where $s$ is the $s$-wave
real component of the gap, as required by group theory for the
in-plane Cu–O lattice symmetry of orthorhombic cuprate
superconductors such as YBCO \cite{tsueirmp}. Although a $d+s$ pair
state in YBCO is established
\cite{polturakprb93,basovprl95,limonovprl98,luprl01,engelhardtprb99},
reports of the magnitude of the $s$-wave admixture \index{$s$-wave
admixture} vary. Twinning of the YBCO thin films used in the
experiments discussed above prevents the determination of possible
real admixtures. To further examine the \index{$d_{x^2-y^2}$-wave
pairing symmetry} $d$-wave order parameter symmetry, it is therefore
important to perform \index{pairing symmetry} pairing symmetry tests
as a function of in-plane momentum using untwinned films. We have
performed such experiments on untwinned optimally-doped YBCO films
based on YBCO/Nb contacts \cite{smildeprl05,kirtleynatphys06}.

\begin{figure}[h!]
\centerline{\includegraphics[width=4.4in]{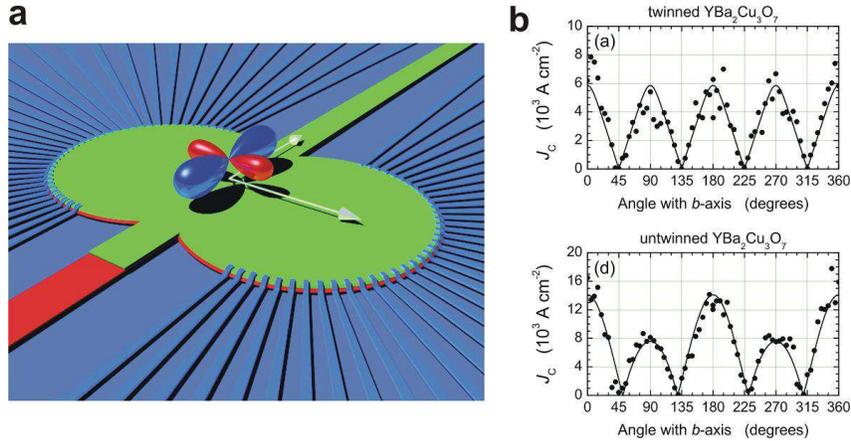}}
\caption[]{(Color) (a)~Angle-resolved electron tunneling with
YBa$_2$Cu$_3$O$_7$/Au/Nb ramp-type junctions oriented every
$5^\circ$ over $360^\circ$. The arrows indicate the main crystal
orientations in the $ab$ plane of the high-$T_c$ superconducting
material. (b)~Critical current densities $J_c$ as a function of the
junction orientation with respect to the YBa$_2$Cu$_3$O$_7$ crystal
for (top) twinned and (bottom) untwinned YBa$_2$Cu$_3$O$_7$ at
$T=4.2$~K and in zero magnetic field [from \cite{smildeprl05}].}
\label{fanangle}
\end{figure}

The experimental layout is summarized in Fig.~\ref{fanangle}a.
Basically, the YBCO base electrode is patterned into a nearly
circular polygon, changing the orientation from side to side by
$5^\circ$. A Au barrier and Nb counterelectrode contact each side.
In this way, the angle with respect to the (010)-orientation is
varied as a single parameter. Figure \ref{fanangle}b presents the
electrical characterization of the twinned base-electrode sample
(top), and the untwinned one (bottom). The superconducting
properties of the Au/Nb bilayer are independent of the orientation.
Therefore, $J_c$ depends on the in-plane orientation $\theta$ with
respect to the $b$-axis of the YBCO crystal only, and presents four
maxima for both samples, approaching zero in between. This is in
agreement with predominant \index{$d_{x^2-y^2}$-wave pairing
symmetry} $d_{x^2-y^2}$-wave symmetry of the superconducting wave
function in one electrode only. In closer detail, the nodes of the
untwinned YBCO sample are found at $5^\circ$ from the diagonal
between the $a$- and the $b$-axis. This presents direct evidence for
a significant real isotropic \index{$s$-wave admixture} $s$-wave
admixture. An estimate for the $s$- over \index{$d_{x^2-y^2}$-wave
pairing symmetry} $d_{x^2-y^2}$-wave gap ratio is $17\%$ for a node
angle $\theta_0=50^\circ$, resulting in a gap amplitude $50\%$
higher in the $b$ (Cu-O chain) direction than in the $a$ direction.
For the twinned base electrode, the nodes are found at the diagonal,
which is expected if all twin orientations are equally present, and
contributions of subdominant components average to zero.

\subsection{Angle-resolved phase-sensitive experiments}

Measurements of the critical currents of single YBCO-Nb junctions
described above resulted in a gap $50\%$ larger in the $b$ than the
$a$ direction for optimally doped YBCO. To further quantify the
deviations from a pure $d_{x^2-y^2}$ symmetry, in particular BTRS
states, \index{phase-sensitive experiment} phase-sensitive
experiments as a function of in-plane momentum are needed. Such
experiments have been suggested to be sensitive to an imaginary
component to the order parameter \cite{beasleyprb94,ngprb04}. We
have performed phase-sensitive
experiments based on two-junction rings connecting YBCO and Nb to be
able to accurately determine the in-plane \index{pairing symmetry}
pairing symmetry \cite{kirtleynatphys06}.

\begin{figure}[h!]
\centerline{\includegraphics[width=4.4in]{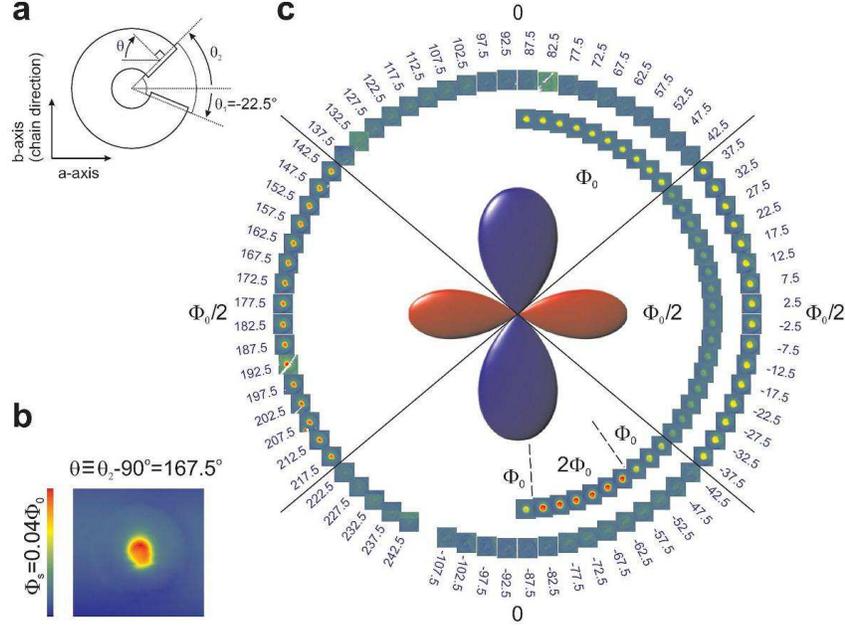}}
\caption[]{(color) (a) Schematic of one ring, with the angles
defined. (b) An image of the ring with second junction normal angle
$\theta=167.5^\circ$ relative to the majority twin a-axis direction,
cooled and imaged in nominally zero field. (c) Images for all of the
rings labelled by $\theta$ and arranged in a polar plot. Rings
cooled in zero field were either in the $n=0$ or the $n=1/2$ flux
quantum states (outer circle); the rings cooled in $0.2~\muµ$T were
in the $n=1/2$, $n=1$ or $n=2$ states (inner circle) [from
\cite{kirtleynatphys06}].} \label{varma}
\end{figure}

A schematic of the rings for this experiment is shown in
Fig.~\ref{varma}a. Each ring has one junction with a fixed angle of
$-22.5^\circ$ relative to the majority twin $a$-axis direction in
the YBCO, with the second junction angle varying in intervals of
$5^\circ$. The ring geometries were optimized with several
considerations in mind. For this measurement it is desirable to have
large rings, because this leads to large $I_cL$ products, so that
the spontaneous magnetization is not reduced from the asymptotic
$\Phi_0/2$ value. Large rings are also easier to resolve with the
SQUID microscope. It is important that the junction interfaces are
well-defined, clean and have a sufficiently high critical current
density.

Figure~\ref{varma} shows a SQUID microscope image of one of the
rings, cooled and imaged in zero field. The spontaneous flux is
clearly visible in the center of the ring; the ring walls are nearly
invisible. Figure~\ref{varma}b shows images of the rings in this
sample, cooled and imaged in zero field, depicted as a polar plot.
This underlines the nearly four-fold symmetry of the data, with the
transitions between $0$-rings and \index{$\pi$-ring} $\pi$-rings
occurring at angle of the variable junction normal $\theta$ values
close to $(2m+1)45^\circ$, $m$ being an integer. The deviations from
four-fold symmetry are systematic and are consistent with the gap
being larger in the $b$-axis direction than in the $a$-axis
direction. The outer circle of images are of the sample cooled in
zero field, in which case the rings either have zero spontaneous
flux (fluxoid number $n=0$) or spontaneous flux close to
$\Phi_0/2~(n =1/2)$. To test whether all of the rings had
sufficiently large junction critical currents to sustain an
appreciable circulating supercurrent, we recooled the sample in a
field of $0.2~\mu$T, with the resulting images shown in the inner
semi-circle of Fig.~\ref{varma}c. In this case the 0-rings were
either in the $n=1$ or 2 state, whereas the \index{$\pi$-ring}
$\pi$-rings remained in the $n=1/2$ state. This shows that the nodal
direction in YBCO films with a predominant twin orientation is
shifted by at least a few degrees from the $(2m+1)45^\circ$ angles
expected for a pure $d_{x^2-y^2}$ superconductor. Furthermore, the
spontaneous fluxes which do not deviate from $\Phi_0/2$ or zero
underline the fact that BTRS is not associated with high temperature
superconductivity in optimally doped YBCO.

\section{Coupling of half-flux quanta}
\label{fluxquanta}

Using grain boundary junctions, it has only been possible to
controllably generate individual \index{$\pi$-ring} $\pi$-ring
\cite{tsueiybcoprl94,tsueirmp}. Such ring has a doubly degenerate
ground state in zero applied flux. This is of interest as model
systems for studying magnetic phenomena $-$including frustration
effects$-$ in Ising antiferromagnets
\cite{aeppliscience97,moessnerprb01,chandraprb88,davidovicprb97,pannetierprl84,lerchprb90}.
Furthermore, studies of coupled $\pi$-loops can be useful for
designing quantum computers based on flux-qubits
\cite{mooijscience99,vanderwalscience00,friedmannature00,ioffenature,blaispra00}
with viable quantum error correction capabilities
\cite{preskillprsl98,bennetnature00}. However, these require a large
number of rings. In this section we concentrate on the realization of
large-scale coupled \index{$\pi$-ring} $\pi$-loop arrays based on
YBCO/Nb Josephson contacts. Scanning SQUID microscopy
\cite{kirtleyibm95,kirtleyapl95} has been used to study the ordering
of \index{half-flux quantum} half-flux quanta in these structures.

\begin{figure}[h!]
\centerline{\includegraphics[width=4.2in]{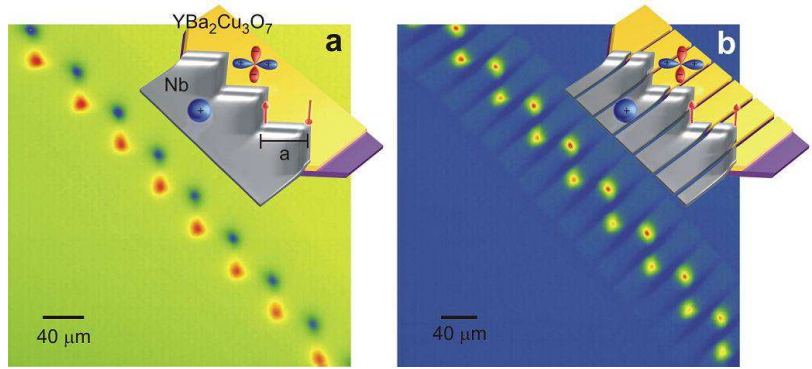}}
\caption[]{(color) Scanning SQUID microscope images of (a) continous
zigzag junction with $40~\mu$m facet lengths, and (b) electrically
disconnected zigzag junction with $40~\mu$m between facet corners
[from \cite{hilgenkampnature}].} \label{arrays1d}
\end{figure}

We have first investigated the generation and coupling of half-integer
flux quanta in the \index{zigzag junction} zigzag array that have been
discussed in Sec. \ref{zigzag}, shown schematically in Fig
\ref{arrays1d} insets.  In the scanning SQUID microscopy image
presented in Fig.  \ref{arrays1d}a, a spontaneously induced magnetic
flux is clearly seen at every corner of the zigzag structure. For this
sample $a=40~\mu$m, which implies that the facets are well within the
wide limit. The observed corner fluxons are arranged in an
antiferromagnetic fashion. This antiferromagnetic ordering was found
to be very robust, occurring for many cool-downs and for different
samples with comparable geometries. Deviations from an
antiferromagnetic arrangement were only observed when a magnetic field
was applied during cool-down, or when an Abrikosov vortex was found
trapped in (or near) the junction interface.

In the zigzag configuration, all the half-fluxons \index{half-flux
quantum} are generated in a singly connected superconducting
structure; the question therefore arises as to whether the
antiferromagnetic ordering is due to a magnetic interaction between
the fractional fluxons, or to an interaction via the superconducting
connection between the corners. To investigate this, we have also
fabricated arrays of corner junctions, in a similar configuration as
the zigzag arrays but with $2.5~\mu$m-wide
slits etched halfway between the corners, as schematically shown in
Fig. \ref{arrays1d}b inset. In this situation there is no
superconducting connection between the separate flux-generating corner
junctions. For a distance between the corners equal to the facet
length in the connected array, $a=40~\mu$m, a ferromagnetic
arrangement of the fractional flux quanta was observed
(Fig. \ref{arrays1d}b). The magnetic interaction between the
 half-flux quanta at these distances is
expected to be very weak, and alignment along minute spurious
background fields in the scanning SQUID microscope is anticipated to
be the dominating mechanism for their parallel arrangement. When the
distance was decreased to about $20~\mu$m, with a slit width of
$1.5~\mu$m, a tendency towards an antiferromagnetic coupling was
observed.

\begin{figure}[h!]
\centerline{\includegraphics[width=3.2in]{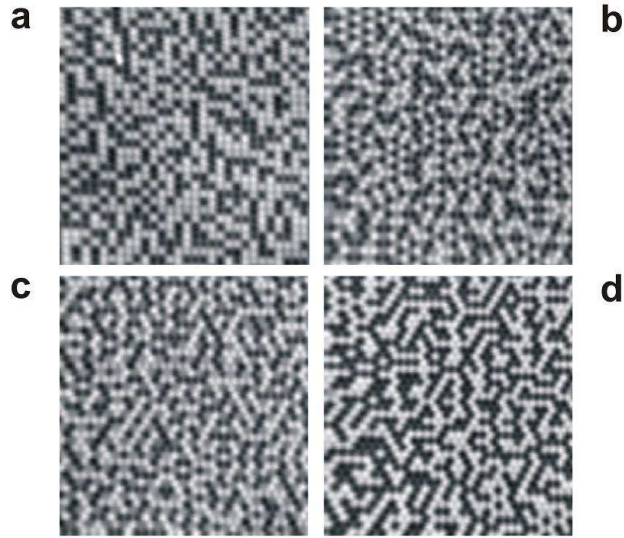}}
\caption[]{SQUID microscopy images of four electrically disconnected
arrays of $\pi$-rings with $2.7~\mu$m junctions and $11.5~\mu$m ring
to ring spacings for (a) square, (b) honeycomb, (c) kagom\'{e}, and
(d) triangle lattices [from \cite{kirtleyprb05}].} \label{arrays2d}
\end{figure}

We also have investigated 2-dimensional \index{$\pi$-ring}
$\pi$-ring arrays made up of individual rings with various
nearest-neighbor distances ($25~\mu$m or closer)
\cite{hilgenkampnature,kirtleyprb05}. The rings were arranged into
arrays with 4 different geometries: square, honeycomb, triangular,
and kagom\'{e}. The square and honeycomb arrays are geometrically
unfrustrated, as their magnetic moments can be arranged so that all
nearest neighbors have opposite spins, and the ground state of these
lattices are only doubly degenerate. In contrast, the triangle and
kagom\'{e} lattices are geometrically frustrated, since it is
impossible for all of the rings to have all nearest neighbors
anti-ferromagnetically aligned, and the ground states are highly
degenerate. Figure \ref{arrays2d} shows examples of scanning SQUID
microscope images of the arrays after cooling in nominally zero
field. Although regions of antiferromagnetic ordering are seen in
the unfrustrated arrays Figs. \ref{arrays2d}a and \ref{arrays2d}b,
anti-ferromagnetic ordering beyond a few lattice distances was never
observed. Nevertheless, antiferromagnetic correlations were seen in
all the 2D \index{$\pi$-ring} $\pi$-ring arrays.

We have shown that it is possible to realize large arrays of
photolithographically patterned  $\pi$-rings.
Half-fluxon Josephson vortices in
electrically connected \index{zigzag junction} zigzag junctions
order with strongly anti-parallel half-fluxon \index{half-flux
quantum} vortices through the superconducting order parameter phase.
Electrically isolated 1D and 2D arrays order much less strongly. The
2D $\pi$-ring arrays show stronger
anti-ferromagnetic correlations than reported previously for 0-ring
arrays \cite{davidovicprl96,davidovicprb97}, but do not order beyond
a few lattice constants. One possibility is that our arrays
correspond to a spin glass \cite{kirtleyprb05}. However, in the
absence of some hidden symmetry breaking, it appears that our rings
are simply doubly degenerate. Furthermore, qualitatively similar results
are obtained for arrays with and without geometrical frustration.
Our experiments with repeated cooling show that there is little
fixed disorder in our arrays. Therefore, it seems unlikely that we
have a spin glass, unless there is some form of frozen-in disorder
that varies from cooldown to cooldown.

\section{Possible applications of $\pi$-phase-shifting elements}

Besides holding a clue to the mechanism of high-$T_c$
superconductivity, the unconventional \index{$d_{x^2-y^2}$-wave
pairing symmetry} $d$-wave symmetry in high-$T_c$ cuprates provides
unique possibilities to realize superconducting
(quantum)-electronics that exploit the associated sign-changes in
the order parameter for orthogonal directions in $k$-space. An
intriguing consequence of this sign-change is the spontaneous
generation of fractional magnetic flux quanta in superconducting
rings incorporating a $d$-wave induced \index{phase-shift}
$\pi$-phase shift. In the following, we discuss some possible
applications of $\pi$-phase-shifting elements for novel
(quantum)-electronics based on \index{high-$T_c$/low-$T_c$ Josephson
contact} high-$T_c$ versus low-$T_c$ Josephson contacts.

\subsection{Complementary Josephson electronics}

\begin{figure}[h!]
\centerline{\includegraphics[width=4.8in]{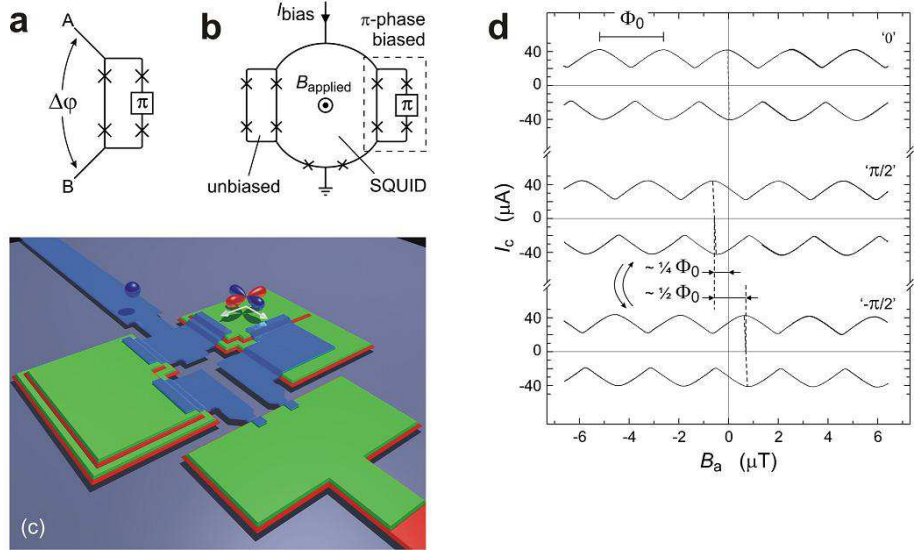}}
\caption[]{(Color) Schematic of (a) a $\pi/2$ phase-shifting element
consisting of four Josephson junctions and a built-in $\pi$ phase
shift, and (b) a switchable $\pi/2$ SQUID, in which a SQUID loop
incorporates the $\pi/2$ phase shifter (dashed rectangle) and an
unbiased four-junction loop. (c) Implementation of the switchable
$\pi/2$ SQUID. (d) Critical current versus applied magnetic field
for a reference SQUID incorporating two unbiased four-junction loops
(top), and for the two complementary states of the switchable
$\pi/2$ SQUID (middle and bottom) [from \cite{smildeaplshifter}].}
\label{bistable}
\end{figure}

For a built-in phase shift of $k\pi$, the $I_c(B)$ characteristics
of $\pi$-SQUIDs are shifted by $\frac{1}{2}k\Phi_0$ compared to a
standard SQUID. Two $I_c(B)$ dependencies are possible, differing
with the polarity of the built-in phase shift. By constructing a
phase-shifting element in which the polarity can be switched, a
bistable superconducting device can be realized, with SQUID
characteristics shifted by $+\frac{1}{2}k\Phi_0$ or
$-\frac{1}{2}k\Phi_0$ compared to the standard case. This
bistability can be used to construct superconducting memory
elements, like flip-flops or programmable logic. Furthermore, Ioffe
et al. (1999) proposed to design qubits for quantum computation
based on the energetically degenerate ground states. They suggested
a possible practical realization with $k=\pm\frac{1}{2}$ (Fig.
\ref{bistable}a). It is based on a superconducting ring containing
four identical Josephson junctions. In this ring a
 $\pi-$phase shift is incorporated using one of
the concepts described earlier. It invokes a persistent circulating
supercurrent $I_{circ}$ relating to a phase drop of practically
$\frac{1}{4}\pi$ per junction, with a slight deviation proportional
to the magnetic flux in the ring, which is equal to the product of
$I_{circ}$ and the inductance of the phase-shifting element
$L_{shifter}$.

We have realized a superconducting bistable device by
incorporating \index{phase-shift} a $\pi/2$ phase-shifting element
into a dc SQUID. Incorporating this element in a larger SQUID
structure, as shown in Fig. \ref{bistable}b, a phase shift over the
terminals A and B of $\pm\pi/2$ is obtained whose polarity
depends on the direction of $I_{circ}$. With the $I_c$'s of the
junctions in the overall SQUID being much smaller than those in the
phase shifter, the final device is bistable, with the $I_c(B)$
characteristics shifted by $+\frac{1}{4}\Phi_0$ or
$-\frac{1}{4}\Phi_0$ compared to a standard SQUID, as shown in Fig.
\ref{bistable}d. Notice that the characteristics of these two states
are complementary to each other, which makes these structures
suitable for the construction of (programmable) complementary
Josephon electronics.

\subsection{Half-flux quanta as information carriers}

\begin{figure}[h!]
\centerline{\includegraphics[width=3in]{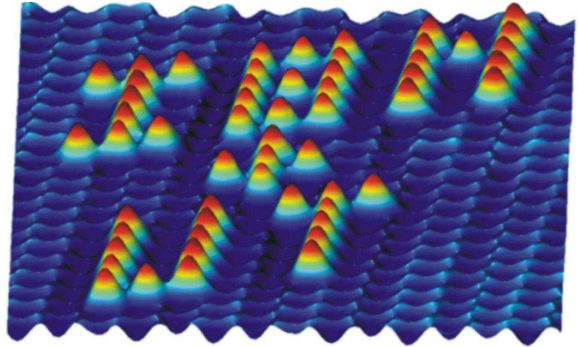}}
\caption[]{Scanning SQUID microscopy image of an array of
half-integer flux quanta forming the characters 'IBM+UT'.}
\label{ibmut}
\end{figure}

The doubly degenerate groundstates of \index{$\pi$-ring} $\pi$-rings
allow their use for information storage. As previously discussed,
under specific circumstances these groundstates are characterized by
the spontaneous formation of half a
quantum of magnetic flux. Their polarity can be set by the action of
a logic (quantum-)gate, making them potentially useful elements for
a random access memory. This is demonstrated in Fig.~\ref{ibmut},
showing a scanning SQUID microscopy image of an array of
half-integer flux quanta forming the
characters 'IBM+UT'. The fluxes were all set to have the same
polarity by cooling in a modest externally applied magnetic field.
The polarity of selected elements was reversed by applying pulses of
control current to a single turn coil incorporated into the SQUID
measurement chip.

\subsection{$\pi$-shifts in Rapid Single Flux Quantum-logic}

In isolated \index{$\pi$-ring} $\pi$-rings, fabricated with
high-$T_c$ grain boundaries \cite{tsueiybcoprl94} or with
connections between \index{high-$T_c$/low-$T_c$ Josephson contact}
high-$T_c$ and low-$T_c$ superconductors \cite{hilgenkampnature},
the generation and manipulation of fractional flux quanta has
already been demonstrated using scanning SQUID microscopy. An
important step towards its application in electronic circuitry is
the incorporation of such $\pi$-loops in
superconducting logic gates, in which a controlled operation on an
electronically applied input signal leads to a predefined output
signal~\cite{beasleyieee,ustinovapl03}. We have performed
experiments realizing this idea \cite{ortleppscience06}, and showed
the first realization of a \index{Toggle Flip-Flop} Toggle Flip-Flop
(TFF) based on Josephson contacts between high-$T_c$ and
low-$T_c$ superconductors, in which the polarity of the fractional
flux quantum provides the internal memory.

The spontaneous generation of fractional flux in the $\pi$-shift
device eliminates the need for the asymmetrically injected bias
current, which reduces the amount of connections to external
control-electronics and allows for symmetry in the design
parameters. This greatly benefits the design-process and
fabrication and also leads to denser circuitry; our first
realization needed only a quarter of the size of a standard Toggle
Flip-Flop in established Niobium technology with the same feature
size of 2.5~$\mu$m.

\subsection{$\pi$-shift qubit concepts}

Over the last decades quantum-computation has received much attention
because of its potential to solve mathematical problems,
which classical computers cannot solve in acceptable times
\cite{divincenzoscience95,divincenzosupmic98}. As usual, it is a long way
from  basic principles to their practical implementation. A
central aspect in this context is the availability of appropriate qubit
elements, which allow integration into logic gates. A qubit is a
quantum-system which, similar to a bit in a conventional computer,
is characterized by two states but which can exist in arbitrary
quantum-superpositions of these states.

While techniques from quantum-optics (cold trapped atoms, photons in
cavities) or from molecular physics (nuclear magnetic resonance
methods) appear to provide promising technologies for the
realization of individual and few qubits, freedom in the
design-parameters of the qubits and upscaling to real computing
devices will require the scalability and variability of solid-state
implementations. Superconductors appear to provide the best starting
point to achieve this goal.

Proposed superconducting qubits operate either with charge
\cite{schnirmanprl97,averinssc98,bouchiatpscripta98,nakamuranature99},
or flux-/phase-states
\cite{bockoieee97,mooijscience99,ioffenature,feigelmanjltp00,blatterprb}.
In the phase-qubit, the information is mainly expressed through the
phase degree of freedom, with only a minor coupling to the
environment or to other qubits. The phase-state itself carries no
charge and no current, rendering the device electrically and
magnetically robust. A first proposal that was made by the ETH
Z\"urich group for a superconducting phase-qubit involved mesoscopic
junctions combining $s$- and \index{$d_{x^2-y^2}$-wave pairing
symmetry} $d$-wave superconductors, so called SDS'-junctions
\cite{ioffenature}. Subsequently, a more easily realizable scheme was
proposed involving superconducting loops with five Josephson
junctions, four conventional ones and a $\pi$-junction
\cite{blatterprb}. In the design, the quantum degrees of freedom
involve only the conventional junctions, while $\pi$-junction acts
primarily as a \index{phase-shift} phase-shifter. A very appealing
aspect hereby is the fact that the qubits do not require a
flux-bias, as is the case for all-low-$T_c$ concepts.


\acknowledgements We would like to thank A.~Brinkman, M.~Dekkers,
A.~Golubov, S.~Harkema, T.~Ortlepp, and F.~Roesthuis for
discussions. This work was supported by the Dutch Foundation for
Research on Matter (FOM), the Netherlands Organization for
Scientific Research (NWO), the Dutch STW NanoNed programme, and the
European Science Foundation (ESF) PiShift programme.


\bibliographystyle{nato}

\bibliography{references}

\printindex
\end{document}